\documentclass[journal]{IEEEtran}

\usepackage[compress]{cite}
\usepackage{stmaryrd}
\usepackage{mathrsfs}
\usepackage{amsfonts}
\usepackage{amsmath}
\usepackage{cases}
\usepackage{txfonts}
\usepackage{amssymb}
\usepackage{epsfig}

\usepackage{algorithm}
\usepackage{algpseudocode}

\newcommand{\fat}[1]{\mbox{\boldmath$#1$}}

\newtheorem{construction}{Construction}

\newtheorem{guess}{Theorem}
\newtheorem{corollary}{Corollary}
\newtheorem{lemma}{Lemma}
\newtheorem{definition}{Definition}

\newtheorem{remark}{Remark}
\newtheorem{example}{Example}

\newcommand{\myQED}{\mbox{}\hfill{$\Box$}}

\begin{document}

\title{Signature Design of Sparsely Spread CDMA Based on Superposed Constellation Distance Analysis}

\author{Guanghui Song,~\IEEEmembership{Member,~IEEE}, Xianbin Wang,~\IEEEmembership{Senior Member,~IEEE}, and Jun Cheng,~\IEEEmembership{Member,~IEEE}}
\maketitle

\begin{abstract}
Sparsely spread code division multiple access (SCDMA) is a non-orthogonal superposition coding scheme that permits a base station simultaneously communicates with multiple users over a common channel. The detection performance of an SCDMA system is mainly determined by its signature matrix, which should be sparse to guarantee large Euclidean distance for the equivalent signal constellation after spreading and superposition. Good signature matrices that perform well under both belief prorogation and the maximum likelihood detections are designed.  The proposed design applies to several similar well-documented schemes, including trellis code multiple access (TCMA), low density spreading, and superposition modulation systems.
\end{abstract}


\section{Introduction}
\IEEEPARstart{T}{he} future
fifth generation (5G) mobile networks are expected to provide an unprecedented capacity in supporting the rapid growth of mobile data traffic with very limited spectrum resources.  New multiple access technique, i.e., non-orthogonal
multiple access (NOMA), which allow multiple concurrent communications,  has been recognized as one of most efficient solutions to fulfill these requirements \cite{nomaVTC,nomaPIMRC,scma,Taherzadeh2014,Nikopour2014}.
Recently, several non-orthogonal code division multiple access (CDMA) schemes, named sparsely spread code division multiple access (SCDMA) \cite{scdmatanaka}\cite{scdmaguo}, low-density spreading \cite{Hoshyar2008}\cite{lds}, and sparse code multiple access \cite{scma,Taherzadeh2014,Nikopour2014}, have been developed for multiple access channels. All of these techniques rely on sparse signature sequences and  near-optimal joint multi-user belief prorogation (BP) detections on sparse graphs. We collectively call these techniques SCDMA.  It has demonstrated many advantages with respect to the capacity load and detection complexity over the conventional dense CDMA and orthogonal multiple access schemes.

In the downlink of a general SCDMA system, a base station simultaneously communicates with multiple users. Data streams for the multiple users are first spread (encoded) into vectors by multiplying their signature sequences, which are sparse and the elements are usually selected from a given alphabet set. Multiple data streams after spreading are superimposed at the base station and broadcasted to the users over a common channel, i.e., using the common resources such as time and frequency. A multi-user BP detection is performed at each user to recover the data streams.

The performance of SCDMA detection is mainly determined by a signature matrix that consists of all the users' signature sequences as its row vectors. Generally, the signature matrix should have a good sparsity, i.e., without short cycles in the formed factor graph, to achieve a good BP detection performance. Theoretically, if its factor graph has no cycles, the BP detection converges to the maximum likelihood (ML) detection performance \cite{BP}. Moreover, the equivalent signal constellation after spreading and superposition should have large Euclidean distance which ultimately determines the performance bound of ML detection. This motivates us to design the elements in the signature matrix in SCDMA.

Signature design has been investigated for dense spreading in conventional CDMA \cite{web,seqcdma,Alishahi2012}, where an orthogonal or low-correlated sequence set are constructed to maximize an equivalent CDMA channel capacity. The problem becomes more complex for sparse spreading in SCDMA since the design should be implemented under the sparsity constraint of the signature matrix. The problem becomes even more difficult when a two-dimensional modulation scheme is employed as in the scenarios of \cite{scma} and \cite{lds}. Works \cite{tcmaicc} and \cite{tcma} show that a user constellation rotation significantly affects detection performance of multi-user superposition codes. Convolutional code is employed for each user in \cite{tcmaicc} \cite{tcma} and the multiple access scheme is referred to as trellis code multiple access (TCMA), which can be regarded as a spatial case of the scenarios in \cite{scma} and \cite{lds} with unitary spreading length. Work \cite{Harshan2011} considers two-user TCMA and designs the user constellation rotation by maximizing an equivalent channel capacity. Work \cite{lds} considers a general multi-user
 SCDMA with a non-trivial spreading length. For a given regular factor graph structure, \cite{lds} shows that a Latin-rectangular signature matrix significantly outperforms a randomly generated signature matrix due to a large minimum code distance property. However, many open research problems, including how to efficiently find an optimal signature matrix with the maximum minimum code distance for an SCDMA system, how to efficiently estimate the ML detection performance, and how to design signature matrix that works well under both ML and BP detections,  are still yet to be resolved.

In this paper, we consider a general SCDMA system with a two-dimensional quadrature amplitude modulation (QAM) and give a theoretical framework for signature design. We give a formal definition of SCDMA code distance and a distance enumerator analysis to estimate the ML detection performance. For a given factor graph structure of an SCDMA code, we design the optimal signature matrix with the maximum minimum code distance. We construct two SCDMA code families whose factor graphs have very few short cycles. The constructed SCDMA codes outperform the existing codes in terms of both word error rate (WER) performance and detection complexity. Our numerical results show that their BP detections exactly converge to their ML detection performances with few iterations. Simulations for turbo-coded SCDMA systems with variety communication rates are given to verify the validity of our design in more practical applications.

The remainder of the paper is organized as follows. Section~\ref{sec:model} describes
the SCDMA system model and introduces three detection algorithms. Section~\ref{sec:distance}
defines the SCDMA code distance and some properties on code distance are shown.
Section~\ref{sec:design} gives the optimal signature matrix design for SCDMA codes. Section~\ref{sec:const} gives two constructions of code families with
few short cycles in their factor graph and
large minimum code distance. Section~\ref{sec:simulation} gives simulations for our design in both uncoded and turbo-coded SCDMA systems. Section~\ref{sec:conclude} concludes this paper.
\section{SCDMA and  Detections} \label{sec:model}
\subsection{System Model}
\begin{figure}
\includegraphics[width=3.1in]{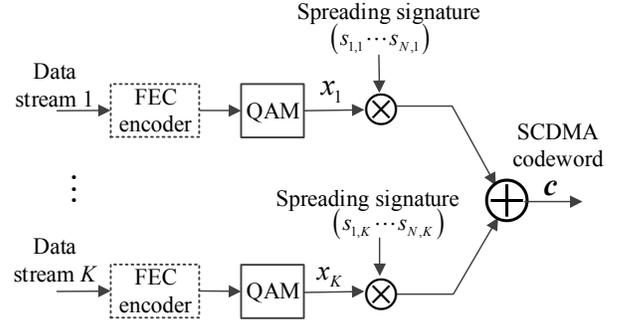}
\centering
\caption{A $K$-user SCDMA transmitter model.} \label{fig:scma}
\end{figure}

Figure~\ref{fig:scma} shows a $K$-user downlink SCDMA transmitter model at the base station. There are $K$ data streams to be transmitted to $K$ mobile users.  After a forward error correction (FEC) encoding, each user's data stream is modulated and spread by multiplying its signature sequence. Figure~\ref{fig:scma} illustrates the spread processing for an individual symbol of each user's data stream.  Here we consider QAM with $x_k\in\mathcal{X}\buildrel \Delta \over
=\{\frac{1}{\sqrt{2}}+\frac{i}{\sqrt{2}}, \frac{1}{\sqrt{2}}-\frac{i}{\sqrt{2}}, -\frac{1}{\sqrt{2}}+\frac{i}{\sqrt{2}}, -\frac{1}{\sqrt{2}}-\frac{i}{\sqrt{2}}\}$, where $i$ is the imaginary unit. The output after spreading is $(s_{1,k},...,s_{N,k})x_k$ for $x_k$, where $(s_{1,k},...,s_{N,k})$ with $s_{n,k}=0$ or $s_{n,k}=e^{i\theta}, \theta\in[0, 2\pi), n=1,...,N$, is called a signature sequence of user $k$. Here we considered unitary energy for each nonzero element of the signature sequence. It should be emphasized that the spreading vector is sparse, i.e., the majority of elements might be 0. Number of nonzero elements in a spreading vector is called an effective spreading length.

The $K$ users' data streams after spreading are superimposed and transmitted over $N$ orthogonal channel resources, e.g., OFDMA tones or MIMO spatial layers. The transmitted vector is represented as
\begin{equation}\label{eq:codeword}
\begin{bmatrix}
          c_1 \\
          c_2 \\
          \vdots \\
          c_N \\
        \end{bmatrix}=\begin{bmatrix}
                        s_{1,1} & s_{1,2} & \cdots & s_{1,K} \\
                      s_{2,1} & s_{2,2} & \cdots & s_{2,K} \\
                        \vdots & \vdots & \ddots & \vdots \\
                      s_{N,1} & s_{N,2} & \cdots & s_{N,K}\\
                      \end{bmatrix}\begin{bmatrix}
          x_1 \\
          x_2 \\
          \vdots \\
          x_K \\
        \end{bmatrix}
\end{equation}
which is referred to as an SCDMA codeword.
Note that there is a total of $4^K$ number of SCDMA codewords corresponding to the $4^K$ different variations of $(x_1,...,x_K)^\textrm{T}$, where $()^\textrm{T}$ is the transpose of a matrix.
Matrix $S=[s_{n,k}]$ is sparse and is referred to as a signature matrix.
By multiplexing $K$ users over $N$ channel resources, the load of the SCDMA code is $K/N$. Since for $K/N\leq1$, we can use orthogonal spreading sequences to achieve near single-user performance, in this paper, we mainly consider overloaded SCDMA with $K/N>1$.

Each user receives
a noise-corrupted codeword $\fat{y}=h\fat{c}+\fat{z}$, where $\fat{c}=(c_1,...,c_N)^\textrm{T}, \fat{y}=(y_1,...,y_N)^\textrm{T}$, $h$ is a channel gain, and $\fat{z}=(z_1,...,z_N)^\textrm{T}$ is a complex Gaussian noise vector with each element an i.i.d. mean-0  variance-$N_0$ complex Gaussian variable, i.e., $z_n\sim\mathcal{CN}(0,N_0)$. A joint $K$-user SCDMA detection is performed to recover the data streams.

\begin{figure}
\includegraphics[width=1.9 in]{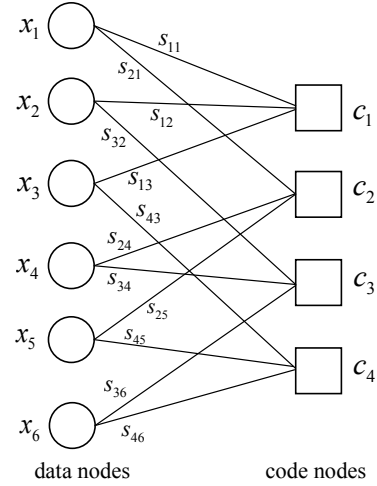}
\centering
\caption{A factor graph representation of a $6$-user $4$-resource SCDMA code.} \label{fig:factor}
\end{figure}

Each SCDMA code can be represented by a sparse factor graph.
Figure~\ref{fig:factor} gives an example of factor graph representation for a $(K=6)$-user $(N=4)$-resource SCDMA code proposed in \cite{scma}\cite{Taherzadeh2014}, where data nodes $x_1,...,x_K$ denote data symbols of $K$ users, and code nodes $c_1,...,c_N$ denote $N$ SCDMA coded symbols.  There is an edge between $c_n$ and $x_k$, denoted as $e_{n,k}$, if $s_{n,k}\neq0$. Let $E$ be the set that includes all the edges in the factor graph. Edge $e_{n,k}\in E$ is labeled by signature element $s_{n,k}$. Each code node is a superposition of its neighboring data nodes, i.e., $c_n=\sum_{\{k|e_{n,k}\in E\}}s_{n,k}x_k$.
The degree of  a node is the number of edges incident with the node. The graph is called code-node regular of degree $d$ if all the the code nodes have degree $d$. Figure~\ref{fig:factor} is code-node regular of degree $3$. If a factor graph is cycle-free, we call it a tree graph, and we call the corresponding code a tree SCDMA code.

\subsection{SCDMA Detections}
In this section, after we briefly reviewing two detection algorithms of SCDMA, the ML and BP detections, we give an approximate BP detection whose detection complexity linearly increases with the user number. For all these three detections, we assume the receiver knows channel gain $h$ perfectly.
\subsubsection{ML detection}
Based on $\fat{y}$, the ML detection is
\begin{equation}\label{eq:ML}
\hat{\fat{c}}=\arg\max_{\fat{c}}\textrm{Pr}(\fat{y}|\fat{c})=\arg\min_{\fat{c}}\|\fat{y}-h\fat{c}\|
\end{equation}
where the ML detection is reduced to the minimum distance detection due to the memoryless Gaussian channel.
If more than one codeword satisfies (\ref{eq:ML}), we randomly select one of them as our decision with equal probability. Here we assume that all SCDMA codewords are transmitted with equal probability, and thus, (\ref{eq:ML}) is equivalent to the maximum a posteriori probability detection.
Based on the estimated SCDMA codeword $\hat{\fat{c}}$, we can uniquely determine the transmitted data stream. The complexity of ML detection is $O(4^K)$ for SCDMA with QAM.
\subsubsection{BP detection}\label{sec:BP}
A suboptimal scheme with lower complexity is BP detection.
This detection is performed on the factor graph. The whole detection is performed iteratively in a belief-propagation manner. In each iteration, each node of the factor graph performs a local processing and exchanges message with its neighboring nodes.

\textbf{Code Node Processing:}
Consider the processing at code node $c_n$. Let $\textrm{P}_{x_k,c_n}^{\ell-1}(x_k=\alpha)$, a priori probability of $x_k=\alpha, \alpha\in\mathcal{X}$, be the message output from data nodes $x_k, k\in\kappa=\{k|e_{n,k}\in E\}$ to code node $c_n$ at the $(\ell-1)$-th iteration. Let $\kappa_{\backslash k}$ be the set obtained by delating $k$ from $\kappa$. Based on this priori probability, and the channel observation $y_n$, code node $c_n$ outputs a probability message of
\begin{eqnarray}
\!\!\!\!\!\!\!\!\!\!\!\!&&\textrm{P}_{c_n,x_k}^{\ell}(x_k\!=\!\alpha)\!=\!\!\!\sum_{\alpha_j\in\mathcal{X},j\in\kappa_{\backslash k}}\!\!\textrm{Pr}(y_n|x_j\!=\!\alpha_j,\!j\!\in\!\kappa_{\backslash k},\! x_k\!=\!\alpha)\!\!\prod_{j\in\kappa_{\backslash k}}\!\!\textrm{P}_{x_j,c_n}^{\ell-1}(x_j\!=\!\alpha_j)\nonumber\\
\!\!\!\!\!\!\!\!\!\!\!\!&&=\!\!\frac{1}{\pi N_0}\!\sum_{\alpha_j\in\mathcal{X},j\in\kappa_{\backslash k}}\!\!\exp\left(-\frac{|y_n\!-\!\!\sum_{j\in\kappa_{\backslash k}}\!\!s_{n,j}\alpha_j\!-\!s_{n,k}\alpha|^2}{N_0}\right)\prod_{j\in\kappa_{\backslash k}}\textrm{P}_{x_j,c_n}^{\ell-1}(x_j\!=\!\alpha_j)\nonumber
\end{eqnarray}
to data node $x_k, k\in\kappa$, which is a probability of $x_k=\alpha, \alpha\in\mathcal{X}$.

\textbf{Data Node Processing:}
Data node $x_k$ combines the message obtained from the code nodes in its neighborhood and outputs a probability message of
\begin{equation}
\textrm{P}_{x_k,c_n}^{\ell}(x_k=\alpha)=\prod_{\{j|e_{j,k}\in E,j\neq n\}}\textrm{P}_{c_j,x_k}^{\ell}(x_k=\alpha), \alpha\in\mathcal{X}\nonumber
\end{equation}
to its neighboring code node $c_n, n\in \{n|e_{n,k}\in E\}$.

\textbf{Hard Decision:}
After a fixed number $L$ of iterations, hard decision is made for $x_k$ as
\begin{equation}
\hat{x}_k=\arg\max_{\alpha\in\mathcal{X}}\prod_{\{j|e_{j,k}\in E\}}\textrm{P}_{c_j,x_k}^L(x_k=\alpha).\label{eq:hard}
\end{equation}

The complexity of BP detection is dominated by the complexity of code node processing, whose complexity is $O(4^d)$, where $d$ is the maximum code node degree. The specific detection complexity is determined by the code node degree profile and the iteration number that is required for the detection.

Note that if the factor graph of an SCDMA code is a tree graph, its BP detection will converge to its ML detection  with finite number of iterations. Many works show that short cycles, such as length-4 cycles, significantly degrade the performance of BP detection \cite{lin}. Therefore, in this paper we only consider factor graphs without length-4 cycles.

\subsubsection{Approximate BP Detection}
In the BP detection, the processing at the code node is a MAP processing, where the accurate probability about the estimated data is calculated. In this section, we give an approximate BP detection with a simplified code node processing. We regard the summation of interferences for each data as complex Gaussian, so we only need to track a mean and variance message.

 Take the processing at code node $c_n$ as an example. The associated receive at this node is
\begin{equation}
y_n=\sum_{\{k|e_{n,k}\in E\}}s_{n,k}x_k+z_n=s_{n,k}x_k+\xi_k\label{eq:interfer}
\end{equation}
where $\xi_k=\sum_{j\in\kappa_{\backslash k}}s_{n,j}x_{j}+z_n$ is the equivalent noise for $x_k$. We approximately regards $\xi_k$ as complex Gaussian, i.e., $\xi_k\sim\mathcal{CN}(\mu_k,N_k)$ with
\begin{eqnarray}
\mu_k\!\!\!\!\!\!\!\!\!\!\!&&=\textrm{E}[\xi_k]=\sum_{j\in\kappa_{\backslash k}}s_{n,j}\textrm{E}[x_{j}]=\sum_{j\in\kappa_{\backslash k}}s_{n,j}\sum_{ \alpha\in\mathcal{X}}\textrm{P}_{x_{j},c_n}^{\ell-1}(x_{j}=\alpha)\alpha\nonumber\\
N_k\!\!\!\!\!\!\!\!\!\!\!&&\!=\!\textrm{E}[|\xi_k\!-\!\mu_k|^2]\!=\!\!\sum_{j\in\kappa_{\backslash k}}\!\textrm{E}[|x_{j}\!-\!\textrm{E}[x_j]|^2]\!+\!N_0\!=\!\sum_{j\in\kappa_{\backslash k}}\left(1\!-\!\left|\textrm{E}[x_{j}]\right|^2\right)\!+\!N_0\nonumber\\
\!\!\!\!\!\!\!\!\!\!\!&&=\sum_{j\in\kappa_{\backslash k}}\left(1-\left|\sum_{ \alpha\in\mathcal{X}}\textrm{P}_{x_{j},c_n}^{\ell-1}(x_{j}=\alpha)\alpha\right|^2\right)+N_0\nonumber
\end{eqnarray}
where $\textrm{E}[*]$ takes the expectation of a random variable.
Therefore, code node $c_n$ outputs a probability message of
\begin{equation}
\textrm{P}_{c_n,x_k}^{\ell}(x_k=\alpha)=\frac{1}{\pi N_k}\exp\left(-\frac{|y_n-s_{n,k}\alpha-\mu_k|^2}{N_k}\right), \alpha\in\mathcal{X}\nonumber
\end{equation}
 to data node $x_k, k\in\kappa$.

The approximate BP detection may work well when the code node degree is large or the noise level is high since at these two cases, interference term
$\xi_k$ is more like Gaussian.

The processing complexity of the code node reduces to $O(d)$, where $d$ is the maximum code node degree.
\section{SCDMA Code Distance and Properties} \label{sec:distance}
In this section, we first define an SCDMA code distance and distance enumerator function, which is used to formulate a union bound for ML detection. Some properties about SCDMA code distance enumerator function and the minimum code distance are derived.

\begin{definition}\label{def:d}
Distance between two SCDMA codewords $\fat{c}, \fat{c}^\prime\in\mathcal{C}$ is
\begin{equation}\label{eq:d}
d(\fat{c},\fat{c}^\prime)=||\fat{c}-\fat{c}^\prime||\nonumber
\end{equation}
where $\mathcal{C}$ is the SCDMA code set.\myQED
\end{definition}

\begin{definition}
\begin{equation}\label{eq:distance}
d_{\min}=\min_{\fat{c}, \fat{c}^\prime\in\mathcal{C},\fat{c}\neq \fat{c}^\prime} d(\fat{c},\fat{c}^\prime)
\end{equation}
is called the minimum distance of SCDMA code $\mathcal{C}$.
\myQED
\end{definition}

Applying (\ref{eq:codeword}) to (\ref{eq:distance}), we obtain the following lemma immediately.
\begin{lemma}\label{lem:distance}
Let $\triangle\mathcal{X}
\buildrel \Delta \over
=\{0, \pm\sqrt{2},\pm\sqrt{2}i,\sqrt{2}\pm\sqrt{2}i,-\sqrt{2}\pm\sqrt{2}i\}$ and  $\triangle\mathcal{X}^K$ be the universal set of length $K$ vectors over $\triangle\mathcal{X}$. The minimum distance of the SCDMA code with spreading signature matrix $S$ is
\begin{eqnarray}
d_{\min}(S)=\!\!\!\!\!\!\!\!&&\min_{\fat{u}\in\triangle\mathcal{X}^K, \fat{u}\neq \textbf{0}} F(S,\fat{u})\nonumber\\
\!\!\!\!\!\!\!\!&& F(S,\fat{u})\buildrel \Delta \over=\sqrt{\sum_{n=1}^N\left|\sum_{k=1}^Ks_{n,k}u_k \right|^2}\nonumber
\end{eqnarray}
where $\fat{u}=(u_1,...,u_K)$ and $\textbf{0}=(0, ...,0)$.
\myQED
\end{lemma}

To give a global description of the code distance spectrum of an SCDMA code, we have the following definition.
\begin{definition}\label{eq:Ad}
Distance enumerator function for an SCDMA code with signature matrix $S$ is
\begin{eqnarray}
A(S, Z)=\!\!\!\!\!\!\!\!&&\frac{1}{|\mathcal{C}|}\sum_{\fat{c}\in\mathcal{C}}\sum_{\fat{c}^\prime\in\mathcal{C},\fat{c}^\prime\neq\fat{c}}Z^{d(\fat{c},\fat{c}^\prime)}\nonumber\\
=\!\!\!\!\!\!\!\!&&\frac{1}{|\mathcal{C}|}\sum_{\fat{c}\in\mathcal{C}}\sum_{\fat{c}^\prime\in\mathcal{C},\fat{c}^\prime\neq\fat{c}}Z^{F(S,\fat{c}-\fat{c}^\prime)}\label{eq:enumerator}
\end{eqnarray}
where $Z$ is a dummy variable, and $|\mathcal{C}|=4^K$ for QAM is the cardinality of the code set.\myQED
\end{definition}

Equation (\ref{eq:enumerator}) in fact gives an average distance spectrum for all the codewords in the code set.

The distance enumerator function of SCDMA code can be used to calculate a multi-user union bound developed in \cite{songIT}. It is an WER upper bound for ML detection.

Let $A(S,Z)=\sum_dA(d)Z^d$ be the distance enumerator function of an SCDMA  code with signature matrix $S$, where $A(d)$ can be regarded as the average number of codeword pairs with distance $d$. The WER $P_W$ under ML detection is upper bounded by\\
\emph{Union Bound \cite{songIT}:}
\begin{equation}
P_W\leq A(0)+\sum_{d>0}A(d)Q\left(\frac{d}{\sqrt{2N_0}}\right).\label{eq:union}
\end{equation}
Note that (\ref{eq:union}) has different form as that in \cite{songIT} since the definition of code distance in this work has different form from that in \cite{songIT}.

We give the following properties for the SCDMA code distance enumerator function and minimum code distance.
\begin{lemma}[Row Rotation Invariance]\label{lem:rot1}
\begin{equation}\label{eq:rot1}
A(S^\prime, Z)=A(S, Z)
\end{equation}
holds for $s^\prime_{n,k}=e^{i\theta_n}s_{n,k}, \theta_n\in[0,2\pi), k=1,...,K, n=1,...,N$.\myQED
\end{lemma}
\emph{Proof:}
Equation (\ref{eq:rot1}) holds because
\begin{equation}
F(S^\prime,\fat{u})\!=\!\sqrt{\sum_{n=1}^N\left|\sum_{k=1}^Ke^{i\theta_n}s_{n,k}u_k \right|^2}\!=\!\sqrt{\sum_{n=1}^N\left|\sum_{k=1}^Ks_{n,k}u_k \right|^2}=F(S,\fat{u})\nonumber
\end{equation}
holds for any $\fat{u}\in\Delta\mathcal{X}^K$. \myQED

\begin{lemma}[Column Rotation Invariance]\label{lem:rot2}
\begin{equation}\label{eq:rot2}
A(S^*, Z)=A(S, Z)
\end{equation}
holds for any $s^*_{n,k}=e^{im_k\pi/2}s_{n,k}, m_k\in Z, k=1,...,K, n=1,...,N$, where $Z$ is the set of integer numbers.\myQED
\end{lemma}

\emph{Proof:}
Equation (\ref{eq:rot2}) holds because
\begin{equation}
F(S^*,\fat{u})\!=\!\sqrt{\sum_{n=1}^N\left|\sum_{k=1}^Ke^{i\frac{m_k\pi}{2}}s_{n,k}u_k \right|^2}\!=\!\sqrt{\sum_{n=1}^N\left|\sum_{k=1}^Ks_{n,k}u^*_k \right|^2}\!=\!F(S,\fat{u}^*)\nonumber
\end{equation}
where $u^*_k=e^{i\frac{m_k\pi}{2}}u_k\in\Delta\mathcal{X}$, holds for any $\fat{u}\in\Delta\mathcal{X}^K$.
\myQED

\begin{lemma}[Add a User or Resource]\label{lem:add}
For a give signature matrix $S$, it holds that
\begin{eqnarray}
d_{\min}(\bar{S}_c)\leq d_{\min}(S)\leq d_{\min}(\bar{S}_r)\nonumber
\end{eqnarray}
where $\bar{S}_r$ and $\bar{S}_c$ are signature matrices obtained by adding a row (resource) and column (user)
to $S$, respectively.
\myQED
\end{lemma}

Similarly, we can obtain an opposite proposition of Lemma~\ref{lem:add} by deleting a user or resource.

\begin{corollary}\label{col:bound}
For a give signature matrix $S$,
\begin{eqnarray}\label{eq:bound}
d_{\min}(S)\leq \sqrt{2w}
\end{eqnarray}
where $w$ is the minimum effective spreading length.
\myQED
\end{corollary}

\emph{Proof:}
Equation (\ref{eq:bound}) is from the fact that  $\sqrt{2w}$ is the minimum distance that is achieved by the matrix obtained by delating all the columns of $S$ except the one with the minimum effective spreading length.
\myQED

\begin{lemma}[Concatenation of Signature Matrices]
\begin{eqnarray}
d_{\min}(S)\geq\sqrt{\sum_{j=1}^n{d_{\min}(S_j)}^2}\nonumber
\end{eqnarray}
where $S=[S_1^T, S_2^T, \cdots, S_n^T]^T$ is a concatenation of $S_j, j=1,...,n$.
\myQED
\end{lemma}

\section{Optimal Signature Matrix}\label{sec:design}
For a given factor graph structure, we design the optimal signature matrix with the maximum minimum SCDMA code distance.

For factor graph $G$, we have infinite number of SCDMA codes by varying its edge labels, i.e., the phases of nonzero elements of the signature matrix.
Let $\mathcal{S}_G$ be the universal set that includes all the possible signature matrices, edge labels, associated with $G$. We aim to find the SCDMA code with the maximum minimum code distance.

Since a disconnected factor graph can be considered as multiple independent SCDMA codes, in the following we only consider connected factor graphs.

\begin{definition}
\begin{equation}
S^\textrm{opt}=\arg\max_{S\in\mathcal{S}_G}d_{\min}(S)\nonumber
\end{equation}
is called an optimal signature matrix of $G$. \myQED
\end{definition}
For a factor graph, there are infinite number of  optimal signature matrices due to Lemmas~\ref{lem:rot1} and \ref{lem:rot2}.

For a given factor graph $G$, finding $S^\textrm{opt}$ in $\mathcal{S}_G$ is a non-convex problem with high complexity.
However, if $G$ is cycle-free, the problem can be simplified. We first give the following Theorem.
\begin{guess}\label{thm:tree}
If $G$ is a cycle-free factor graph, for each signature matrix $S\in\mathcal{S}_G$, there exists a matrix  $S^*\in\mathcal{S}_G^*\!=\!\left\{S\left|s_{n,k}\!=\!e^{i\theta_k} , \theta_1\!=\!0, \theta_2,...,\theta_K\!\in\![0, \frac{\pi}{2}), \textrm{for}\ e_{n,k}\in E\right. \right\}$ with
$A(S^*,Z)=A(S,Z)$.
\myQED
\end{guess}

\emph{Proof:}
We first prove that for each $S\!\in\!\mathcal{S}_G$ there exists $S^\prime\in\mathcal{S}_G^\prime=\left\{S\left|s_{n,k}\!=\!e^{i\theta_k}, \theta_1\!=\!0, \theta_2,...,\theta_K\!\in\!(-\infty, \infty), \textrm{for}\ \ e_{n,k}\in E\right. \right\}$ with $A(S^\prime,Z)=A(S,Z)$.
Using the row rotation invariance property of Lemma~\ref{lem:rot1}, we just need to show that for a given $S\!\in\!\mathcal{S}_G$, there exists an $S^\prime\!\in\!\mathcal{S}^\prime_G$ which is a row rotation of $S$. Assume that $S$ is given. We determine $S^\prime$ as follows. Since zero elements in $S^\prime$  are predetermined by the factor graph $G$, we only determine the nonzero elements in $S^\prime$ in the following steps:

i.  For each $n\in\{n|e_{n,1}\in E\}$,  the $n$-th row
of $S^\prime$ is a rotation

\  \ \ of the $n$-th row of $S$, i.e., $s^\prime_{n,k}=s_{n,k}/s_{n,1}$ for $e_{n,k}\in E$.

ii.  Find a column of $S^\prime$ that has only one determined nonzero

\  \ \ element and at least one undetermined element. Assume

\ \ \ that the $m$-th column is found and the only determined

\  \ \  nonzero element is $s^\prime_{j,m}$.  For each $n\in\{n|e_{n,m}\in E, n\neq j\}$,

\ \ \ the $n$-th row of $S^\prime$ is a rotation of the $n$-th row of $S$, i.e.,

 \ \ \ $s^\prime_{n,k}=s^\prime_{j,m}s_{n,k}/s_{n,m}$ for $e_{n,k}\in E$.

iii. If all the elements of $S^\prime$ are determined, terminate the

\ \ \ \ procedure, otherwise, repeat step ii.\\
We show that Step ii can always be successful if there exists undetermined elements in $S^\prime$. First, if there exists undetermined element in $S^\prime$, we can always find a column with both determined and undetermined nonzero elements since $G$ is connected and the nonzero elements are determined in a row-by-row manner according to the procedure. Moreover, if  a column has both determined and undetermined nonzero elements,  the number of determined nonzero element must be one. Suppose that there are more than one  determined nonzero elements in column $m$ in Step ii, i.e., elements $s^\prime_{j,m}\neq0$ and $s^\prime_{j^\prime,m}\neq0, j^\prime\neq j$, are determined. Since the labeling begins from the edges incident with data node $x_1$ (in Step i), there should exist two paths from data node $x_1$ to both code nodes $c_j$ and $c_{j^\prime}$, i.e., there exists a path between $c_j$ and $c_{j^\prime}$ that goes through $x_1$. Since there exists another path of $c_j\rightarrow x_m\rightarrow c_{j^\prime}$ between $c_j$ and $c_{j^\prime}$, which results a cycle in $G$. This conflicts with the fact that  $G$ is cycle-free. Therefore, Step ii can always be successful.

It also holds that $S^\prime\!\in\!\mathcal{S}^\prime_G$, since
Step i guarantees that the nonzero elements in the first column are 1, and Step ii guarantees that nonzero elements in each column are the same. Note that each column except the first with more than one nonzero elements will be found in Step ii, otherwise, we can show that a cycle exists in the graph similarly. Since $S^\prime$ is a row rotation of $S$, $A(S^\prime,Z)=A(S,Z)$.

Direct applying the column rotation invariance property of Lemma~\ref{lem:rot2}, we can get a matrix $S^*\in\mathcal{S}_G^*$  through column rotations from $S^\prime$ with $A(S^*,Z)=A(S^\prime,Z)=A(S,Z)$. Thus, the theorem is proved.
\myQED

\begin{corollary}\label{cor:tree}
If $G$ is a cycle-free factor graph, there exists an optimal signature matrix with
$S^\textrm{opt}\in\mathcal{S}_G^*$. \myQED
\end{corollary}

We consider a special tree factor graph with $K$ data nodes and one code node with load $K$ in which case the signature matrix become a vector. This is the scenario considered in TCMA \cite{tcmaicc}\cite{tcma}.
We first consider the simplest case of $K=2$.

\begin{figure}
\includegraphics[width=3.3 in]{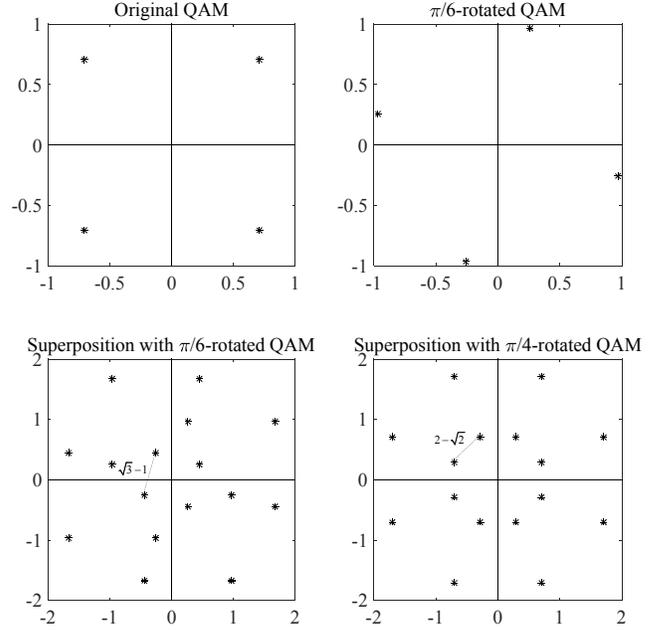}
\centering
\caption{Constellation diagram of two SCDMA code sets for $K=2, N=1$. One is for the optimal signature matrix $[1,\ e^{i\pi/6}]$ and the other is for signature matrix $[1,\ e^{i\pi/4}]$ used in \cite{tcmaicc}\cite{tcma}.} \label{fig:constellation}
\end{figure}

\begin{table}
\caption{Optimal signature matrix and minimum distance for $N=1, K\leq6$.}
\label{tab:vector}
\begin{center}
\begin{tabular}{c|c|c}
\hline\hline
$K$& Optimal signature vector & $\delta_K$\\
\hline
1& [1] \ \ \ \ \ \ \ \ \ \ \ \ \ \  \ \ \ \ \ \ \ \ \ \ \ \ \ \ \ \ \ \ \ \ \ \ \ \ \ \ \ \ \ \ \ \ \ \ \  & $\sqrt{2}$\\
2& $[1\  e^{i\pi/6}]$\ \ \ \ \ \ \ \ \ \ \ \ \ \ \ \ \ \ \ \ \ \ \ \ \ \ \ \ \ \ \ \ \ \ \ \ \ \ \ \ \ \ \ \  & $\sqrt{3}-1$ \\
3& $[1\  e^{i0.0974\pi}\ e^{i0.4026\pi}]$ \ \ \ \ \ \ \ \ \ \ \ \ \ \ \ \ \ \ \ \ \ \ \ \ \ \ \ \ \  & $0.4310$\\
4& $[1\  e^{i0.0477\pi}\ e^{i0.0947\pi}\ e^{i0.1965\pi}]$\ \ \ \ \ \ \ \ \ \ \ \ \ \ \ \ \ \ \ \ & $0.2086$ \\
5& $[1\  e^{i0.0851\pi}\ e^{i0.1368\pi}\ e^{i0.1631\pi}\ e^{i0.1894\pi}]$\ \ \ \ \ \ \ \ \ \  & $0.1142$\\
6& $[1\  e^{i0.0266\pi}\ e^{i0.0664\pi}\ e^{i0.1696\pi}\ e^{i0.473\pi}\ e^{i0.4866\pi}]$ & $0.0595$\\
\hline\hline
\end{tabular}
\end{center}
\end{table}

\begin{guess}\label{thm:two-user}
For the factor graph of an SCDMA system with $K=2$ users sharing $N=1$ resource, $S^\textrm{opt}=[1,\ e^{-i\pi/6}]$ is an optimal signature matrix with $d_{\min}(S^\textrm{opt})=\sqrt{3}-1$.
\myQED
\end{guess}

\emph{Proof:} See Appendix~\ref{app:two-user}. \myQED

Using (\ref{eq:enumerator}), the distance enumerator function of the two-user SCDMA code with the optimal signature matrix is calculated as: $2Z^{\sqrt{3}-1}+\frac{1}{4}Z^{\sqrt{6}-\sqrt{2}}+5Z^{\sqrt{2}}+\frac{9}{4}Z^2+Z^{\sqrt{8-2\sqrt{3}}}+Z^{\sqrt{6}}+2Z^{\sqrt{3}+1}+Z^{\sqrt{8+2\sqrt{3}}}+\frac{1}{4}Z^{2\sqrt{3}}+\frac{1}{4}Z^{2\sqrt{2+\sqrt{3}}}$,
which could be used to estimate the WER performance based on the union bound (\ref{eq:union}).

Signature matrix $[1,\ e^{i\pi/4}]$ is used in \cite{tcmaicc}\cite{tcma} for two-user TCMA system. It is suboptimal since it gives a smaller minimum distance of $2-\sqrt{2}$. We illustrated the superimposed constellation for both cases in Fig.~\ref{fig:constellation}.

It is difficult to formulate the optimal signature matrix for $N=1, K>2$. We obtain the optimal signature matrix and minimum distance, denoted as $\delta_K$, for $K\leq6$ in TABLE~\ref{tab:vector} by numerical search. The minimum distance decreases as the number of users $K$ increases.

In general, a signature matrix of a multi-resource SCDMA system with each row an optimal signature vector for a single-resource SCDMA system may not be optimal.
We only have the following lower bound for the minimum code distance of such signature matrix.
\begin{definition}
Let $\alpha\subset \mathfrak{N}\buildrel \Delta \over
=\{1, 2, \cdots, N\}$ be an index subset. Subgraph $G_{\setminus\alpha}$ is obtained by delating all the data nodes that are adjacent to the code nodes with index in $\alpha$ and edges induced from these data nodes. \myQED
\end{definition}

\begin{lemma}\label{lem:bound}
Let factor graph $G$ be code-node regular degree of $q, q>1$.
Let $S\in\mathcal{S}_G$ be a signature matrix with each row an optimal signature vector of a single-resource SCDMA system. It holds that
\begin{eqnarray}
d_{\min}(S)\geq\min_{\alpha\subset \mathfrak{N}}\sqrt{n_1(G_{\setminus\alpha}){\delta_1}^2+n_2(G_{\setminus\alpha}){\delta_{q}}^2}\nonumber
\end{eqnarray}
where $n_1(G_{\setminus\alpha})$ and $n_2(G_{\setminus\alpha})$ are the numbers of code nodes in $G_{\setminus\alpha}$ with degree one and larger than one, respectively.
\myQED
\end{lemma}

\emph{Proof:} See Appendix~\ref{app:bound}. \myQED

Note that Lemma~\ref{lem:bound} can be extended to the case of irregular code-node degree factor graph with a more complex expression.

\begin{example}[An Optimal $K$-User, $(K-1)$-Resource Tree SCDMA Code]\label{eg:tree}
We have the following $K$-user $(K-1)$-resource tree SCDMA code with each row an optimal two-user single-resource signature vector,
\begin{equation}\label{eq:tree1}
S^\textrm{opt}_{K-1,K}=\begin{bmatrix}
                        1 & e^{\pi/6} & & & \\
                         & e^{\pi/6}  &1 & &  \\
                        &  & 1& e^{\pi/6}& \\
                         &  & & e^{\pi/6}&1\\
                          &  & & &\cdots\\
                      \end{bmatrix}.\nonumber
 \end{equation}
 This SCDMA code has a load of $K/(K-1)$. Using Lemma~\ref{lem:bound} we have $d_{\min}(S^\textrm{opt}_{K-1,K})\geq\min\{\sqrt{K-1}(\sqrt{3}-1),\sqrt{2}\}$.  Also based on the proof of Theorem~\ref{thm:two-user} in Appendix~\ref{app:two-user}, we know that $d_{\min}(S^\textrm{opt}_{K-1,K})=\min\{\sqrt{K-1}(\sqrt{3}-1),\sqrt{2}\}$ exactly holds. For $K\geq5$, the minimum distance is $\sqrt{2}$ which achieves the upper bound of Corollary~\ref{col:bound}. In fact, we can further show that $S^\textrm{opt}_{K-1,K}$ is also an optimal signature labeling for $K<5$.
\myQED
\end{example}

For a factor graph with cycles, we can find an edge set so that after delating the edges in the set, the remaining graph is a tree. We simplify its labeling based on the following theorem.
\begin{guess}\label{thm:nontree}
Let $G$ be a factor graph with cycles and $\varphi\subset E$ be an edge subset that after delating the edges $\varphi$, the remaining graph is a tree. For each signature matrix $S\in\mathcal{S}_G$, there exists a matrix  $S^*\in\!\mathcal{S}_G^*(\varphi)=\{S|s_{n,k}\!=\!e^{i\theta_k} , \theta_1\!=\!0, \theta_2,...,\theta_K\!\in\![0, \frac{\pi}{2}), \textrm{for}\ e_{n,k}
\in \bar{\varphi},\textrm{ and }s_{n,k}\!=\!e^{i\theta_{n,k}}, \theta_{n,k}\!\in\![0, 2\pi), \textrm{for}\ e_{n,k}
\in \varphi\}$ with
$A(S^*,Z)=A(S,Z)$.
\myQED
\end{guess}

\emph{Proof:}
See Appendix~\ref{app:nontree}.
\myQED

\begin{corollary}\label{cor:nontree}
For factor graph $G$ with cycles and $\varphi$ defined in Theorem~\ref{thm:nontree}, there exists an optimal signature matrix with $S^\textrm{opt}\in\mathcal{S}_G^*(\varphi)$. \myQED
\end{corollary}

\begin{example}[Optimal Labeling for Factor Graph in Fig.~\ref{fig:factor}] \label{eg:optmat}
Based on Corollary~\ref{cor:nontree}, we can simplify the labeling for the $6$-user and $4$-resource SCDMA factor graph in Fig.~\ref{fig:factor}
proposed in \cite{scma}\cite{Taherzadeh2014} as
\begin{equation}
S_{4,6}=\begin{bmatrix}
                        1 & e^{i\theta_2} & e^{i\theta_3}&0 &0&0 \\
                         1&0 &0&e^{i\theta_4}&e^{i\theta_5}&0  \\
                        0&e^{i\theta_2} &0 &e^{i\theta_{3,4}}&0&e^{i\theta_6} \\
                         0& 0&e^{i\theta_3}& 0&e^{i\theta_{4,5}}&e^{i\theta_{4,6}}\\
                      \end{bmatrix}\label{eq:optmat}
 \end{equation}
 where $\theta_j\in[0,\pi/2), j=2,...,6$ and $\theta_{3,4}, \theta_{4,5}, \theta_{4,6}\in[0,2\pi)$
 since after delating the edges in set $\varphi=\{e_{3,4}, e_{4,5}, e_{4,6}\}$, the factor graph will be a tree. Its load is $1.5$. We can find 4 length-6 cycles in its factor graph.
 Through a full search based on (\ref{eq:optmat}), we obtain the following optimal signature matrix
 \begin{equation}
S_{4,6}^{\textrm{opt}}=\begin{bmatrix}
                        1 & e^{i0.1431\pi} & e^{i0.2021\pi}&0 &0&0 \\
                         1&0 &0&e^{i0.3127\pi} &e^{i0.3765\pi}&0  \\
                        0&e^{i0.1431\pi}&0 &e^{i0.5736\pi}&0&e^{i0.2667\pi} \\
                         0& 0&e^{i0.2021\pi}& 0&e^{i0.3935\pi}&e^{i0.3078\pi}\\
                      \end{bmatrix}\nonumber
 \end{equation}
 which has the minimum distance $d_{\min}(S_{4,6}^{\textrm{opt}})=1.3726$. Note that the Latin-rectangular labeling proposed in \cite{scma}\cite{lds}, which uses the elements $\{1, e^{i\pi/6}, e^{i\pi/3}\}$ with permutation for each row, only gives the minimum distance of $1.1658$.
\myQED
\end{example}

\section{Two Constructions of Code Families}\label{sec:const}

In this section, we give two constructions of code families whose factor graph has very few short cycles.  We give examples to obtain the optimal signature labeling for these two constructions.

\begin{construction}[A $Kq$-User, $(K-1)q$-Resource SCDMA Code Family]\label{const:improve}

\begin{equation}\label{eq:tree3}
S_{(K-1)q,Kq}=\begin{bmatrix}
                        I & \fat{v}_1 I & & & &\fat{v}_{K-1} P \\
                         & \fat{v}_1 I & \fat{v}_2 I & & & \\
                         &  & \fat{v}_2 I&\ddots& & \\
                         &  & &\ddots & \fat{v}_{K-2} I & \\
                          &  & && \fat{v}_{K-2} I& \fat{v}_K I\\
                      \end{bmatrix}\nonumber
 \end{equation}
where $\fat{v}_k=[e^{i\theta^k_1} \cdots e^{i\theta^k_q} ], \theta^1_j,\cdots ,\theta^{K-1}_j\in [0, \pi/2)$ for $j=1,...,q$, and $\theta^K_j=\theta^{K-1}_j\in [0, \pi/2)$, for $j=1,...,q-1$, $\theta^K_q\in [0, 2\pi)$, $I$ is a  $q\times q$  identity matrix, and
 $P$ is the following $q\times q$ permutation matrix
 \begin{equation}
P=\begin{bmatrix}
                       0&0&\cdots&0& 1\\
                         1&0&\cdots &0& 0\\
                         0&1&\cdots &0 &0\\
                         \vdots & &\ddots& &\vdots \\
                           0&0&\cdots& 1&0& \\
                      \end{bmatrix}\label{eq:P}.
 \end{equation}
This SCDMA code family has a load of $K/(K-1)$.
Vectors $\fat{v}_j, j=1,...,K$, should be optimized to achieve the maximum minimum distance. It is easy to see that there is only one length-$2(K-1)q$ cycle in its corresponding factor graph. Since if we delete one edge in $\varphi=\{e_{(K-1)q,Kq}\}$ the graph will be cycle-free, we have used the simplified labeling for the remaining tree graph based on Corollary~\ref{cor:nontree}.
\myQED
\end{construction}

\begin{example}[An Optimal $6$-User, $4$-Resource SCDMA Code]\label{eg:plus46}
Consider $K=3, q=2$ in Construction~\ref{const:improve}. The graph will be cycle-free by deleting $\varphi=\{e_{4,6}\}$. Through a full search, we obtain the following optimal $6$-user, $4$-resource SCDMA code:
 \begin{equation}
\bar{S}_{4,6}^{\textrm{opt}}=\begin{bmatrix}
                         1 & 0 & e^{i\pi/6}&0 &0&e^{i\pi/6} \\
                         0& 1&0&e^{i\pi/6} &e^{i\pi/3}&0  \\
                        0&0&e^{i\pi/6}&0&e^{i\pi/3}&0 \\
                         0& 0&0& e^{i\pi/6}&0&-1\\
                      \end{bmatrix}\nonumber
 \end{equation}
 with the minimum distance $d_{\min}(\bar{S}_{4,6}^{\textrm{opt}})=1.2679$.
\myQED
\end{example}

\begin{remark}
If we delete the edges corresponding to $\fat{v}_{K-1} P$ in Construction~\ref{const:improve}, the graph will become a tree, and the maximum minimum code distance will reduce to $\min\{\sqrt{K-1}(\sqrt{3}-1),\sqrt{2}\}$, which is achieved by allocating single-resource optimal signature vector to each row as in Example~\ref{eg:tree}.
Introducing the part of $\fat{v}_{K-1} P$ in Construction~\ref{const:improve} increases the minimum code distance for $K<5$, and thus, improves the performance of ML detection.
\myQED
\end{remark}

\begin{example}[An Optimal $8$-User, $6$-Resource SCDMA Code]\label{eg:plus68}
Similarly, by considering $K=4, q=2$ in Construction~\ref{const:improve}, we obtain the following optimal $8$-user, $6$-resource SCDMA code:
 \begin{equation}
S_{6,8}^{\textrm{opt}}=\begin{bmatrix}
                         1 & 0            & e^{i\pi/6}&0              &0             &0    &0&e^{i\pi/6}        \\
                         0& e^{i\pi/3}&0              &e^{i\pi/6} &0            &0     &e^{i\pi/6}&0         \\
                         0&0              &e^{i\pi/6}&0               &e^{i\pi/3}&0   & 0&0        \\
                         0&0             &0              &e^{i\pi/6}  &0             &e^{i\pi/3} &0\\
                         0&0             &0              &0               &e^{i\pi/3}&0  &e^{i\pi/6}  &0     \\
                         0& 0            &0              &0               &0             &e^{i\pi/3}&0&e^{i\pi/6}\\
                      \end{bmatrix}.\nonumber
 \end{equation}
 Its load is $4/3$ and the minimum distance is $d_{\min}(S_{6,8}^{\textrm{opt}})=\sqrt{2}$, which achieves the upper bound of Corollary~\ref{col:bound}.
\myQED
\end{example}

The following construction gives a higher load SCDMA code family.
\begin{construction}[A $Kq$-User, $(K-2)q$-Resource SCDMA Code Family]\label{const:nontree}
\begin{equation}\label{eq:tree2}
S_{(K-2)q, Kq}=\begin{bmatrix}
                       P_{1,1} &\fat{v}_1P_{1,2} & \fat{v}_2P_{1,3}& & &  \\
                         & \fat{v}_1 P_{2,1} &\fat{w}_1P_{2,2} &\fat{v}_3P_{2,3}& & \\
                         &  &  \fat{v}_2P_{3,1}  & \fat{w}_2P_{3,2} &\fat{v}_4P_{3,3} &\\
                         &  &  & \fat{v}_3P_{4,1} &\fat{w}_3P_{4,2} &\fat{v}_5P_{4,3} \\
                          &    &   & & \cdots& \\
                      \end{bmatrix}\nonumber
 \end{equation}
 where $\fat{v}_k=[e^{i\theta^k_1} \cdots e^{i\theta^k_q} ], \theta^k_j\in [0, \pi/2), k=1,...,K-1$, $\fat{w}_k=[e^{i\tau^k_1} \cdots e^{i\tau^k_q} ], \tau^k_j\in [0, 2\pi), k=1,2...,K-3, j=1,...,q$, and $P_{k,j}, k=1,...,K-2, j=1,2,3$, are $q\times q$ permutation matrices.
 This SCDMA code family has a load of $K/(K-2)$.  Permutation matrices $P_{k,j}$ should be carefully selected to avoid short cycles, and vectors $\fat{v}_k, \fat{w}_k$ should be optimized to achieve the maximum minimum distance. Since if we delete edges corresponding to $\fat{w}_kP_{k+1,2}, k=1,2...,K-3$, in Construction~\ref{const:nontree} the graph will be cycle-free, we have used the simplified labeling for the remaining tree graph.
\myQED
\end{construction}

\begin{example}[An Optimal $8$-User, $4$-Resource SCDMA Code]\label{eg:84}
Consider $K=4, q=2$ in Construction~\ref{const:nontree}. By selecting $P_{1,1}=P_{1,2}=P_{1,3}=P_{2,1}=P_{2,3}=I$ and $P_{2,2}=P$ defined in (\ref{eq:P}), the generated factor graph has only one length-$8$ cycle. Since the graph will be cycle-free by deleting the edge in $\varphi=\{e_{3,6}\}$, using Theorem~\ref{thm:nontree}, we obtain the following optimal signature matrix
\begin{equation}
S_{4,8}^{\textrm{opt}}=\begin{bmatrix}
                        1 & 0 & e^{i\theta_2}&0 &e^{i\theta_4}&0&0&0 \\
                         0& e^{i\theta_1}  &0&e^{i\theta_3}&0&e^{i\theta_5}&0&0  \\
                        0&0&e^{i\theta_2}&0&0&e^{i\theta_{3,6}} &e^{i\theta_6}&0\\
                         0& 0 &0&e^{i\theta_3}&e^{i\theta_4}&0&0&e^{i\theta_7}\\
                      \end{bmatrix}\nonumber
 \end{equation}
 where $(\theta_1,...,\theta_7)=(0.2618\pi, 0.1435\pi, 0.1279\pi, 0.2297\pi,$ $0.3505\pi, 0.3935\pi, 0.361\pi)$ and $\theta_{3,6}=0.2269\pi$. Its load is $2$
and the minimum code distance is $d_{\min}(S_{4,8}^{\textrm{opt}})=0.8305$.
\myQED
\end{example}

\begin{remark}
 If we allocate the single-resource optimal signature vector for each row of $S_{(K-2)q, Kq}$ in Construction~\ref{const:nontree},
using Lemma~\ref{lem:bound}, we can show that
 the minimum distance $d_{\min}(S_{Kq,(K-2)q})\geq\min\{d_3\sqrt{K-2},\sqrt{2}\}$. For $K\geq13$, the minimum distance is $\sqrt{2}$, which achieves the upper bound of Corollary~\ref{col:bound}.\myQED
\end{remark}


%
\begin{figure}[t]
\includegraphics[width=3.3 in]{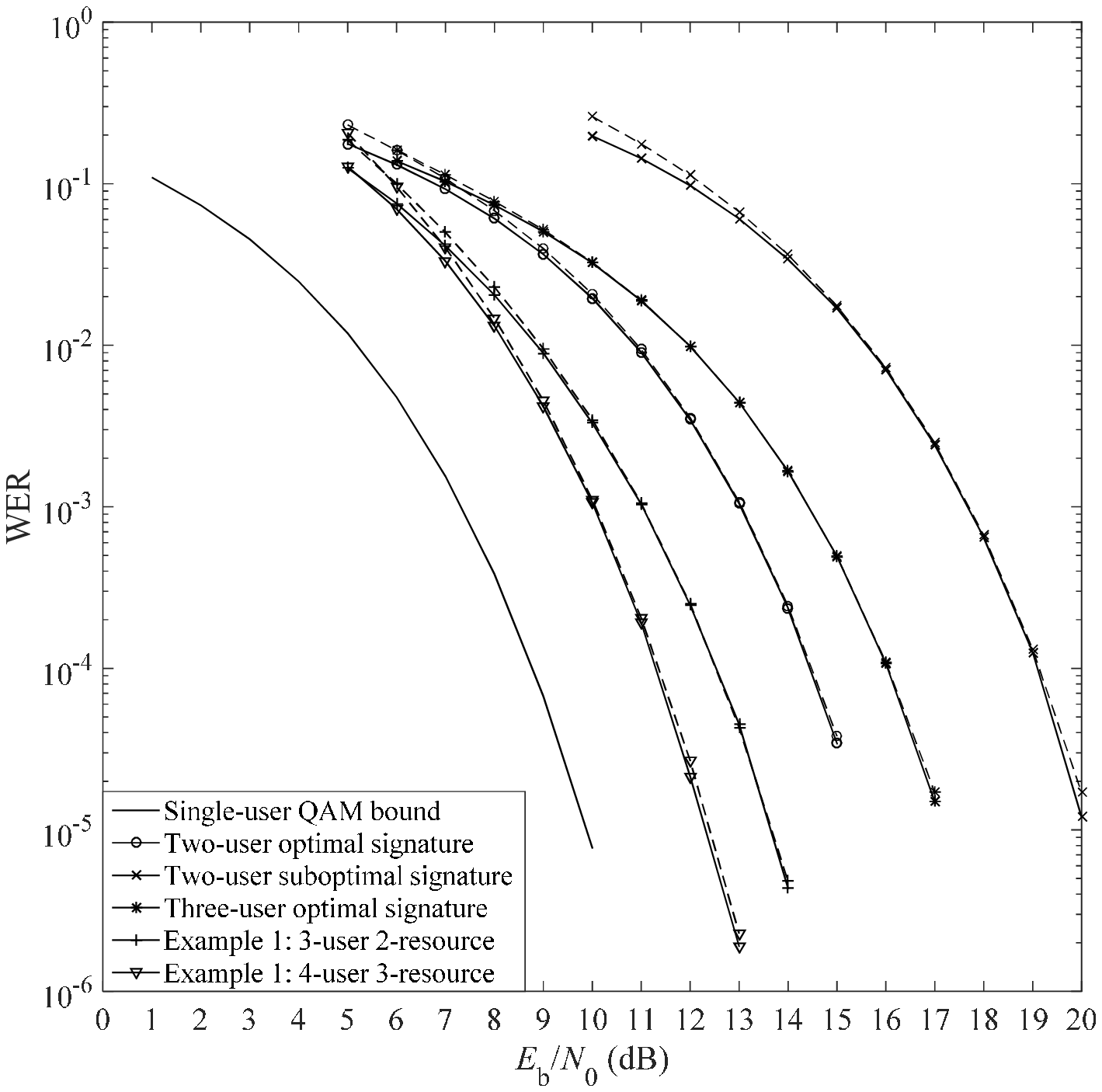}
\centering
\caption{WER of optimal tree SCDMA codes, i.e., two and three-user single-resource optimal codes obtained in TABLE~\ref{tab:vector} and optimal codes  constructed in Example~\ref{eg:tree} with $K=3, 4$, under ML detection and their union bounds. The WER of two-user suboptimal signature $[1, e^{i\pi/4}]$ used in \cite{tcmaicc}\cite{tcma} and its union bound are also illustrated.} \label{fig:singlesource}
\end{figure}
\begin{figure}[t]
\includegraphics[width=3.3 in]{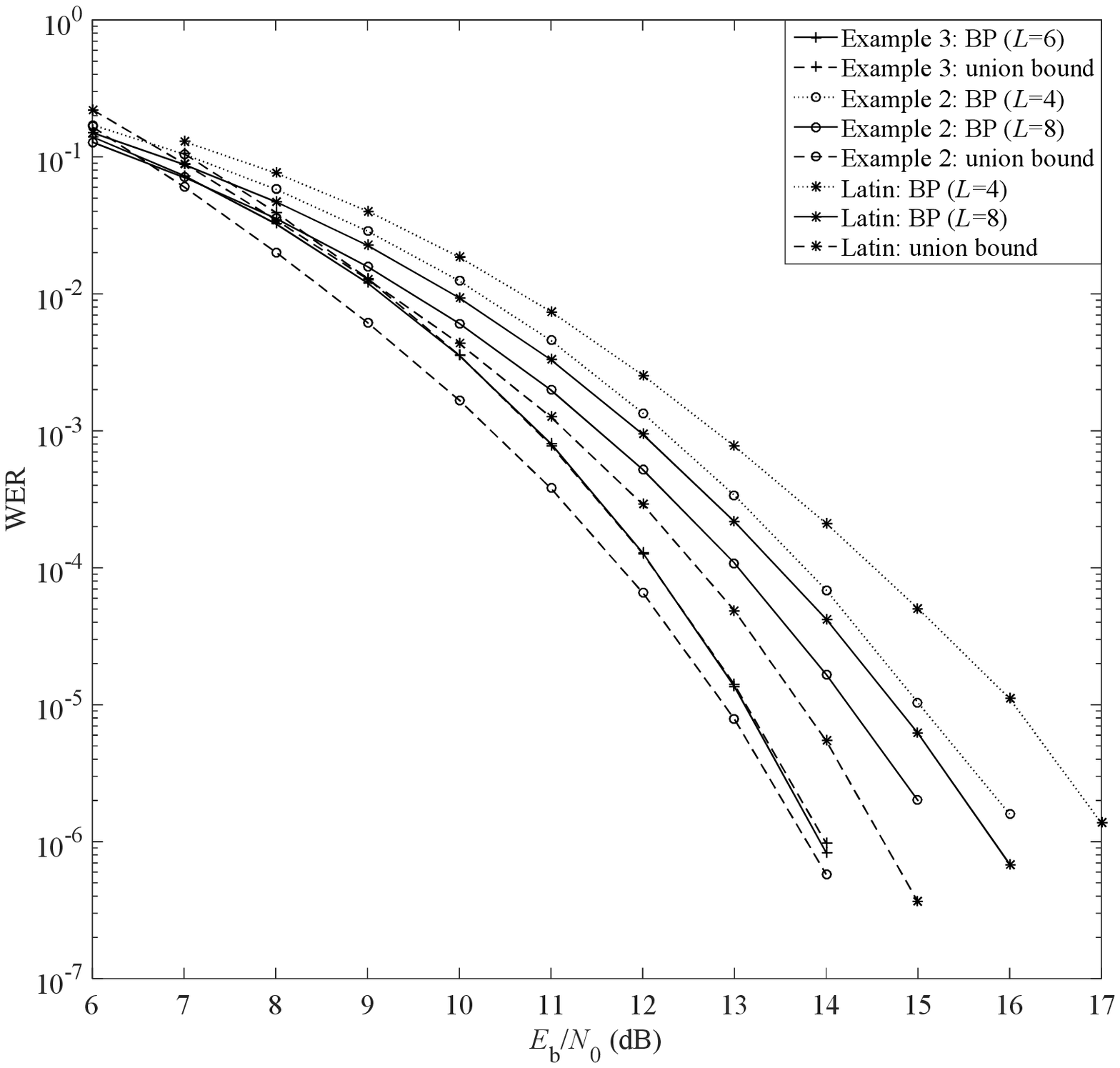}
\centering
\caption{WER of $6$-user $4$-resource SCDMA codes (the optimal codes obtained in Examples~\ref{eg:optmat}, \ref{eg:plus46}, and the code with Latin-rectangular labeling proposed in \cite{scma}\cite{lds}) under BP detections with $L=4, 6, 8$ iterations and their union bounds.} \label{fig:foursix}
\end{figure}

\section{Simulations}\label{sec:simulation}
In this section we give error performance simulations for the (uncoded) SCDMA codes designed in Sections~\ref{sec:design} and \ref{sec:const} and turbo-coded SCDMA over AWGN channel.

\subsection{Uncoded SCDMA}
Figure~\ref{fig:singlesource} illustrates the WER curves (solid lines) and the union bounds (dashed lines) for optimal tree SCDMA codes, i.e., two and three-user single-resource ($N=1$) optimal codes obtained in TABLE~\ref{tab:vector} and optimal codes constructed in Example~\ref{eg:tree} with $K=3, 4$, under ML detections.  For these codes, the ML and BP detections have exactly the same performance. It shows that the code with higher load has higher WER because higher load results in smaller minimum code distances and more error events (larger distance enumerator coefficients). The two-user optimal signature $[0, e^{i\pi/6}]$ has an asymptotic performance gain of near 2 dB over the suboptimal signature of $[0, e^{i\pi/4}]$ used in \cite{tcmaicc}\cite{tcma}. All simulations coincide well with their union bound (\ref{eq:union}) in most area except a little mismatch at the low WER regime. Therefore, the union bound gives a good estimation for the WER of ML detection.

Figure~\ref{fig:foursix} illustrates the WER of $6$-user $4$-resource SCDMA codes: the optimal codes obtained in Examples~\ref{eg:optmat}, \ref{eg:plus46}, and the code with Latin-rectangular labeling proposed in \cite{scma}\cite{lds} under BP detections and their union bounds. The code obtained in Example~\ref{eg:optmat} with the optimal signature has the best union bound since it has the maximum minimum code distance. The union bound is near 1 dB better than that of the code with Latin-rectangular labeling proposed in \cite{scma}\cite{lds}. For the WER simulation under BP detection with $L=4$ and $8$ iterations, the optimal signature has asymptotic performance gains of about $1$ dB and $0.5$ dB over the Latin-rectangular labeling. However, both of their WER under BP detection with $8$ iterations is more than 1 dB worse than their union bounds due to too many short cycles in their factor graph. Although the code designed in Example~\ref{eg:plus46} of Construction~\ref{const:improve} has a slightly worse union bound than that of the code obtained in Example~\ref{eg:optmat}, its BP detection with 6 iterations converges to its union bound, which means that its BP detection may converge to its ML detection. It has a performance gain of more than 1 dB under BP detection over the optimal code obtained in Example~\ref{eg:optmat} with even fewer iterations.

\begin{figure}[t]
\includegraphics[width=3.3 in]{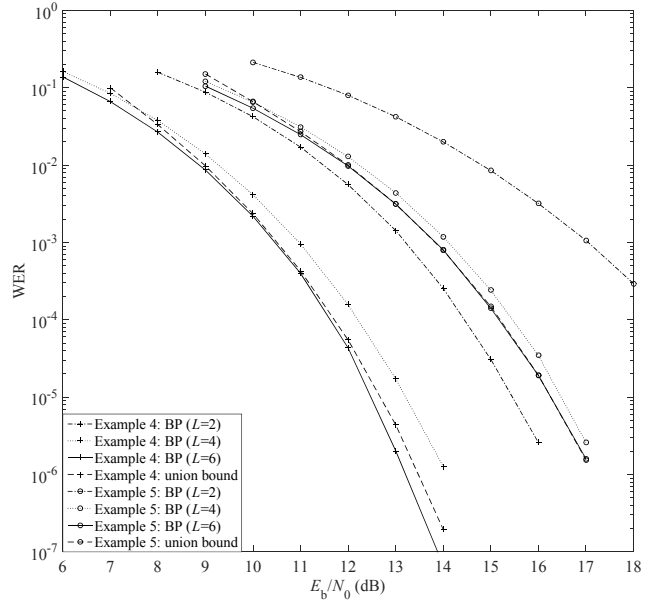}
\centering
\caption{WER of the optimal $8$-user $6$-resource and $8$-user $4$-resource SCDMA codes obtained in Examples~\ref{eg:plus68}, \ref{eg:84} under BP detections with $L=2, 4, 6$ iterations and their union bounds.} \label{fig:eightuser}
\end{figure}

Figure~\ref{fig:eightuser} illustrates the WER of the optimal 8-user 6-resource and 8-user 4-resource SCDMA codes designed in Examples~\ref{eg:plus68} and \ref{eg:84} under BP detections with $L=2, 4, 6$ iterations and their union bounds. With 6 iterations, their BP detections converge to or even exceed their union bounds which means that their BP detections may converge to their ML detections.

Let's consider their BP detection complexities. As mentioned in Section~\ref{sec:BP}, the complexity is manly determined by the code node degrees and the iteration numbers. The factor graph of the code in Example~\ref{eg:optmat} has 4 degree-3 code nodes, and its BP detection requires 8 iterations to converge (according to our simulation observations). The factor graphs of the codes designed in Examples~\ref{eg:plus46}--\ref{eg:84} have 2, 2, and 4 degree-3 code nodes, respectively, and their BP detections only require 6 iterations to converge. Let $C_{\textrm{Eg}.j}$ denote the detection complexity of the code designed in Example $j$. Based on a full consideration of their code node degree profile and iteration number, we can rank their complexities as $C_{\textrm{Eg}.2}>C_{\textrm{Eg}.5}>C_{\textrm{Eg}.4}>C_{\textrm{Eg}.3}$.

\subsection{Turbo-Coded SCDMA}
\begin{figure}[t]
\includegraphics[width=3.3 in]{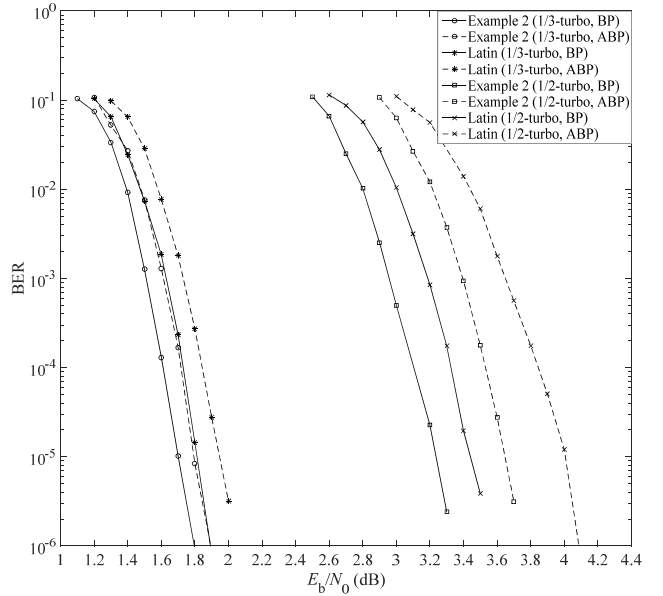}
\centering
\caption{BER of turbo-coded (with turbo coding rates $1/3$ and $1/2$) $6$-user $4$-resource SCDMA codes (the optimal code obtained in Example~\ref{eg:optmat} and the code with Latin-rectangular labeling proposed in \cite{scma}\cite{lds}) under BP and approximate BP (ABP) detections with $30$ iterations.} \label{fig:bercoded1}
\end{figure}
In this section, we give bit-error-rate (BER) simulations of turbo-coded SCDMA systems with QAM modulation, where the FEC code in Fig.~\ref{fig:scma} is realized by a turbo code. Here the turbo code we considered consists two 8-state parallel concatenated convolutional codes with generator matrix $[1, \frac{1+D+D^3}{1+D^2+D^3}]$, which is used in 3GPP LET networks.  By puncturing its parity bits, we can obtain different turbo encoding rates: $1/3, 1/2, 2/3, 4/5$.  For all the simulations, the data stream length for turbo encoding of each user is $1024$. For both BP and approximate BP detections, the global decoding iteration (each global iteration includes a turbo decoding iteration and an SCDMA iteration) number is 30, which is enough for all the considered decodings converge to their best performances.

Moreover, codes given by Example~\ref{eg:tree} and Constructions~\ref{const:improve}, \ref{const:nontree} have irregular effective spreading profile, i.e., effective spreading lengths for symbols of different users may be different. To realize user fairness, we alternately use
column permutations of a signature matrix so that each user's symbol is spread with equal effective spreading length in average sense.
For example, the signature matrix given in Example~\ref{eg:plus46} has effective spreading length profile $(1, 1, 2, 2, 2, 2)$
for the six users. In our simulations,
we divide the modulated symbol stream within a turbo codeword of each user
 into three sub-streams with equal length. The first sub-streams of the six users are spread based on signature matrix $\bar{S}_{4,6}^{\textrm{opt}}$
in Example~\ref{eg:plus46}. For the second and third sub-streams
we use permuted matrices $\bar{S}_{4,6}^{\textrm{opt}}P(1,3)P(2,4)$ and $\bar{S}_{4,6}^{\textrm{opt}}P(1,5)P(2,6)$,
respectively, where $P(i,j)$ is a $6\times6$ column permutation matrix that swaps columns $i$ and $j$.
The permuted signature matrices have the same distance property with the original matrix but
have the effective spreading length profiles $(2, 2, 1, 1, 2, 2)$ and $(2, 2, 2, 2, 1, 1)$, respectively.
By doing this, the average effective spreading length for each symbol, which is the same for each user, becomes $1/3+2\cdot1/3+2\cdot1/3=5/3$.
Therefore, the detection error rate of each user will also be the same.

Figure~\ref{fig:bercoded1} illustrates BER of rate-$1/3$ and $1/2$ turbo-coded $6$-user $4$-resource SCDMA systems under BP and approximate BP (ABP) detections, where the optimal SCDMA code obtained in Example~\ref{eg:optmat} and the code with Latin-rectangular labeling proposed in \cite{scma}\cite{lds} are considered. The sum communication rates of these two turbo-coded SCDMA systems are $1/3\cdot3/2\cdot2=1$ bit/resource and $1/2\cdot3/2\cdot2=3/2$ bit/resource. The rate-$1/3$ turbo-coded SCDMA system with the optimal SCDMA code designed in Example~\ref{eg:optmat} has a performance gain of about $0.1$ dB over the same rate SCDMA system with Latin-rectangular labeling under both BP and ABP decodings. This gain increases if we considered higher rate turbo code, which works at higher $E_\textrm{b}/N_0$ regime, i.e., the gain increases to $0.2\sim0.4$ dB for the rate-$1/2$ turbo-coded SCDMA system. Comparing with BP decoding, the performance loss of the ABP is about $0.1$ dB for rate-$1/3$ turbo-coded SCDMA system at low $E_\textrm{b}/N_0$ regime since the interference term in (\ref{eq:interfer}) is very similar to Gaussian. This performance loss increases to $0.4\sim0.55$ dB at the high $E_\textrm{b}/N_0$ regime for the rate-$1/2$ turbo coded SCDMA system.

\begin{figure}[t]
\includegraphics[width=3.3 in]{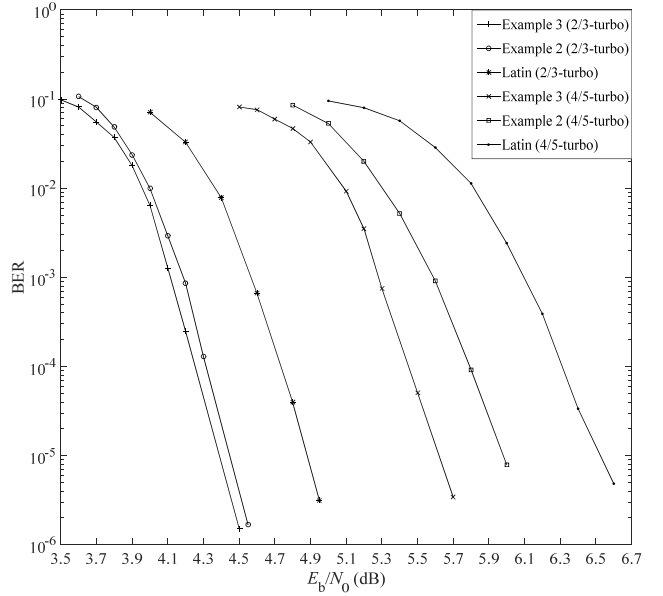}
\centering
\caption{BER of turbo-coded (with turbo coding rates $2/3$ and $4/5$) SCDMA codes (the optimal codes obtained in Examples~\ref{eg:optmat}, \ref{eg:plus46} and the code with Latin-rectangular labeling) under BP detection with $30$ iterations.} \label{fig:bercoded2}
\end{figure}

Figure~\ref{fig:bercoded2} illustrates BER of rate-$2/3$ and $4/5$ turbo-coded SCDMA systems under BP detection, where the optimal codes obtained in Examples~\ref{eg:optmat}, \ref{eg:plus46} and the code with Latin-rectangular labeling are considered. The rate-$2/3$ turbo-coded SCDMA system with the optimal SCDMA codes designed in Examples~\ref{eg:plus46} has slightly better BER than the code designed in Examples~\ref{eg:optmat} and has a performance gain of about $0.5$ dB over the code with the Latin-rectangular labeling. They have the sum communication rate of $2$ bit/resource. For the even higher encoding rate, i.e., a rate-$4/5$ turbo-coded SCDMA system that works at higher $E_\textrm{b}/N_0$ regime, this gain increases and the code in Example~\ref{eg:plus46} has larger performance gains of about $0.4$ dB and $1$ dB over the code in Example~\ref{eg:optmat} and the code with the Latin-rectangular labeling. In this case, the sum communication rate reaches $12/5$ bit/resource.

\begin{figure}[t]
\includegraphics[width=3.3 in]{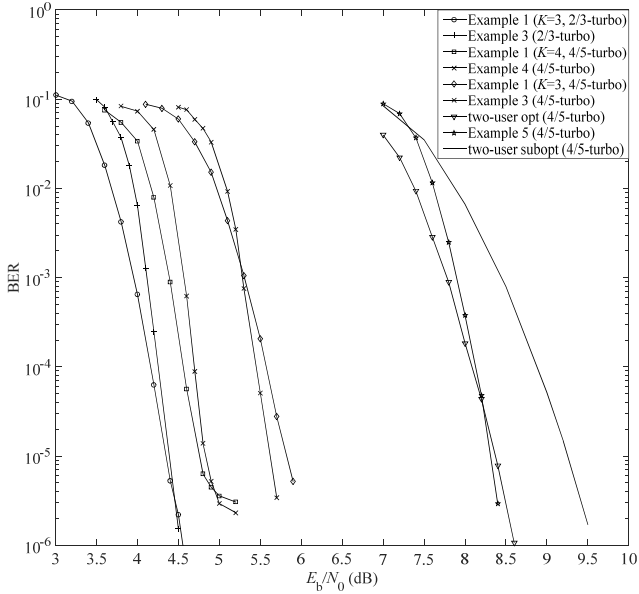}
\centering
\caption{BER of turbo-coded (with turbo coding rates $2/3$ and $4/5$) SCDMA codes (the optimal codes obtained in Examples~\ref{eg:tree}, \ref{eg:plus46}-\ref{eg:84}, and two-user single-resource SCDMA with optimal [Theorem~\ref{thm:two-user}] and suboptimal \cite{tcmaicc}\cite{tcma} labeling) under BP detection with $30$ iterations.} \label{fig:bercoded3}
\end{figure}

Figure~\ref{fig:bercoded3} compares four pairs of turbo-coded SCDMA systems:

a) Rate-$2/3$ turbo-coded SCDMA systems with $(K=3)$-user 2-resource optimal tree SCDMA code constructed in Example~\ref{eg:tree} and $(K=6)$-user 4-resource SCDMA code designed in Example~\ref{eg:plus46}. Their communication rate is $2$ bit/resource.

b) Rate-$4/5$ turbo-coded SCDMA systems with $(K=4)$-user 3-resource optimal tree SCDMA code constructed in Example~\ref{eg:tree} and $(K=8)$-user 6-resource SCDMA code with the optimal SCDMA code designed in Example~\ref{eg:plus68}. Their communication rate is $32/15$ bit/resource.

c) Rate-$4/5$ turbo-coded SCDMA systems with $(K=3)$-user 2-resource optimal tree SCDMA code constructed in Example~\ref{eg:tree} and $(K=6)$-user 4-resource SCDMA code designed in Example~\ref{eg:plus46}. Their communication rate is $12/5$ bit/resource.

d) Rate-$4/5$ turbo-coded SCDMA systems with two-user single-resource optimal SCDMA code obtained in Theorem~\ref{thm:two-user} and $(K=8)$-user 4-resource SCDMA code designed in Example~\ref{eg:84}. Their communication rate is $16/5$ bit/resource.

Each pair has the same communication rate but the code with more users has steeper BER cure, better asymptotic BER performance, due to the joint multi-user processing gain. The rate-$4/5$ turbo-coded two-user single-resource optimal SCDMA code still has a 1 dB performance gain over the same rate turbo-coded suboptimal SCDMA code used in \cite{tcmaicc}\cite{tcma}.

\section{Concluding Remarks} \label{sec:conclude}

We gave a code distance analysis and signature optimization for overloaded SCDMA systems. Good SCDMA codes that work well under both BP and ML detections with low detection complexities are constructed. The constructed codes can support very diverse high-rate services.

As an initial work, we only analyzed the code distance of uncoded SCDMA systems, i.e., without FEC code, and SCDMA with QAM and equal power for each user. One possible extension is to do distance analysis for coded SCDMA systems, which leads to a joint FEC and SCDMA code design. The new system can be treated as a concatenated code. Some works related to concatenated code are given in \cite{songIT,turbo,turbolike}. Another possible extension is to consider a more general modulation and unequal-power user transmissions.

Although we focused on SCDMA systems, our design also apply to several similar well-documented system proposals, such as  TCMA \cite{tcmaicc,tcma,Harshan2011} and superposition modulation \cite{BICM}.
\appendices
\section{Proof of Theorem~\ref{thm:two-user}}\label{app:two-user}
\emph{Proof:} We first prove that there exists an optimal signature matrix in $[1,\ e^{i\theta}], \theta\in[0, \pi/4)$.
According to Corollary~\ref{cor:tree},  optimal signature matrix exists in $[1\ e^{i\theta}], \theta\in[0, \pi/2)$.
Moreover, signature matrices $S=[1, e^{i(\pi/4+\theta)}], \theta\in[0,\pi/4)$ and $S^*=[1, e^{i(\pi/4-\theta)}]$ give the same distance enumerator function since for any $\fat{u}=[u_1,\ u_2]\in\triangle\mathcal{X}^2$,
\begin{eqnarray}
F(S^*,\fat{u})\!\!\!\!\!\!\!\!&&=|u_1+e^{i(\pi/4-\theta)}u_2|=|\overline{u}_1+e^{-i(\pi/4-\theta)}\overline{u}_2|\nonumber\\
\!\!\!\!\!\!\!\!&&=|\overline{u}_1+e^{i(\pi/4+\theta)}\overline{u}_2e^{-i\pi/2}|=F(S,\fat{u}^*)\nonumber
\end{eqnarray}
holds
with $\fat{u}^*=[\overline{u}_1\ \overline{u}_2e^{-i\pi/2}]\in\triangle\mathcal{X}^2$, where $\overline{u}$ is the complex conjugate  of $u$.

To continue prove Theorem~\ref{thm:two-user}, we simplify the expression of minimum distance as
\begin{eqnarray}
\min_{\fat{u}\in\triangle\mathcal{X},\fat{u}\neq \textbf{0}}\left|u_1\!+\!\!s_2u_2\right|\!=\!\min\!\left\{ \min_{u_2\in\triangle\mathcal{X}}|\sqrt{2}\!+\!\!s_2u_2|,\min_{u_2\in\triangle\mathcal{X}}|\sqrt{2}(1\!+\!i)\!+\!\!s_2u_2|\right\}\nonumber
\end{eqnarray}
due to the following facts:
\begin{eqnarray}
 \!\!\!\! \!\!\!\!\!\!\!\!&&\textrm{for}\  u_1=0, \min_{u_2\in\triangle\mathcal{X},u_2\neq0}|u_1\!+\!s_2u_2|=\sqrt{2}\geq \min_{u_2\in\triangle\mathcal{X}}|\sqrt{2}\!+\!s_2u_2|,\nonumber\\
 \!\!\!\! \!\!\!\!\!\!\!\!&& \textrm{for}\  u_1=\pm\sqrt{2}, \pm\sqrt{2}i, \ \min_{u_2\in\triangle\mathcal{X}}|u_1\!+\!s_2u_2|=\min_{u_2\in\triangle\mathcal{X}}|\sqrt{2}\!+\!s_2u_2|, \nonumber\\ \!\!\!\!\!\!\!\!\!\!\!\!&&\textrm{for}\ u_1\!=\!\pm\sqrt{2}(1\!+\!i),
 \pm\sqrt{2}(1\!-\!i), \nonumber\\ \!\!\!\!\!\!\!\!\!\!\!\!&&\ \ \ \ \ \ \ \ \ \min_{u_2\in\triangle\mathcal{X}}|u_1\!+\!s_2u_2|\!=\!\min_{u_2\in\triangle\mathcal{X}}|\sqrt{2}(1\!+\!i)\!+\!s_2u_2|. \nonumber
 \end{eqnarray}

 Since for any $s_2=e^{i\theta}, \theta\in[0, \pi/4)$, the following holds
\begin{eqnarray}
\!\!\!\!\!\!\!\!\!\!\!\!&&\min_{u_2\in\triangle\mathcal{X}}|\sqrt{2}\!+\!s_2u_2|=\sqrt{2}\min\left\{|1-s_2|, |1+(-1+i)s_2|\right\}, \nonumber\\
\!\!\!\!\!\!\!\!\!\!\!\!&&\min_{u_2\in\triangle\mathcal{X}}|\sqrt{2}\!+\!\!\!\sqrt{2}i\!+\!s_2u_2|=\sqrt{2}\min\left\{|1\!+\!i\!-s_2|, |1\!+\!i\!-(1+i)s_2|\right\},\nonumber\\
\!\!\!\!\!\!\!\!\!\!\!\!&&|1-s_2|\leq |1+i-(1+i)s_2|,\nonumber\\
\!\!\!\!\!\!\!\!\!\!\!\!&&|1+(-1+i)s_2|=|-s_2\overline{(1+(-1+i)s_2)}|=|1\!+\!i\!-s_2|\nonumber
\end{eqnarray}
we obtain the final expression of minimum distance as
\begin{equation}
d_{\min}(S)=\sqrt{2}\min\left\{|1-s_2|,  |1\!+\!i\!-s_2|\right\}.\nonumber
\end{equation}
Since for $s_2=e^{i\theta}, \theta\in[0, \pi/4)$, $|1-s_2|$ increases as $\theta$ increases, and $|1+i-s_2|$ decreases as $\theta$ increases,
the optimal $s_2$ should satisfy  $|1-s_2|=|1\!+\!i\!-s_2|$, which leads to $s_2=e^{i\pi/6}$, i.e., $S^\textrm{opt}=[1,\ e^{i\pi/6}]$ with $d_{\min}(S^\textrm{opt})=\sqrt{3}-1$. The theorem is proved. \myQED

\section{Proof of Lemma~\ref{lem:bound}}\label{app:bound}

\emph{Proof:} Let $\beta\subseteq\mathcal{K}\buildrel \Delta \over
=\{1,...,K\}$ be an index subset. Let $\mathcal{U}(\beta)=\{\fat{u}|\fat{u}\in \Delta\mathcal{X}^K, u_k\neq0 \textrm{ for } k\in\beta, u_k=0 \textrm{ for } k\in\bar{\beta}\}$, where $\bar{\beta}$ is the complementary set of $\beta$.
From Lemma~\ref{lem:distance},
\begin{eqnarray}
d_{\min}(S)\!\!\!\!\!\!\!\!\!\!\!\!&&=\min_{\beta\subseteq \mathcal{K},\beta\neq\phi}\min_{\fat{u}\in\mathcal{U}(\beta)} F(S,\fat{u})\nonumber\\
\!\!\!\!\!\!\!\!\!\!\!\!&&=\min_{\beta\subseteq \mathcal{K},\beta\neq\phi}\min_{\fat{u}\in\mathcal{U}(\beta)}\sqrt{\sum_{n\in\gamma}\left|\sum_{k\in\beta}s_{n,k}u_k \right|^2}\nonumber
\end{eqnarray}
where $\gamma=\{n|n\in \mathfrak{N}, s_{k,n}\neq0\textrm{ for some } k\in\beta\}$. Since each row of $S$ is a length-$q$ optimal signature vector for a single-resource SCDMA system, for a given $n\in\gamma$,
 \begin{eqnarray}
 \left|\sum_{k\in\beta}s_{n,k}u_k \right|\geq1_{\{|\beta\cap\{k|s_{n,k}\neq0\}|=1\}}\delta_1+1_{\{|\beta\cap\{k|s_{n,k}\neq0\}|>1\}}\delta_q\nonumber
  \end{eqnarray}
 hold for any $\fat{u}\in\mathcal{U}(\beta)$, where $1_{\{E\}}=1$ if $E$ holds, otherwise, $1_{\{E\}}=0$. Since $\delta_1>\delta_q$, $\left|\sum_{k\in\beta}s_{n,k}u_k \right|$ decreases as $|\beta\cap\{k|s_{n,k}\neq0\}|$ increases. Moreover, since $|\beta\cap\{k|s_{n,k}\neq0\}|\leq d_ n$,  where $d_ n$ is the degree of the $n$-th code node in $G_{\backslash \bar{\gamma}}$, we have
   \begin{eqnarray}
  \left|\sum_{k\in\beta}s_{n,k}u_k \right|\geq1_{\{d_ n=1\}}\delta_1+1_{\{d_ n>1\}}\delta_q\nonumber\\
\sqrt{\sum_{n\in\gamma}\left|\sum_{k\in\beta}s_{n,k}u_k \right|^2}\geq\sqrt{n_1(G_{\setminus\bar{\gamma}}){\delta_1}^2+n_2(G_{\setminus\bar{\gamma}}){\delta_{q}}^2}\nonumber
     \end{eqnarray}
for any  $\fat{u}\in\mathcal{U}(\beta)$. Since by varying $\beta$, $\bar{\gamma}$ can be any proper subset of $\mathfrak{N}$, by denoting $\alpha=\bar{\gamma}$
we obtain
     \begin{eqnarray}
 \min_{\beta\subseteq \mathcal{K},\beta\neq\phi}\min_{\fat{u}\in\mathcal{U}(\beta)}\!\!\sqrt{\sum_{n\in\gamma}\left|\sum_{k\in\beta}s_{n,k}u_k \right|^2}\!\!\geq\!\min_{\alpha\subset \mathfrak{N}}\!\sqrt{n_1(G_{\setminus\alpha}){\delta_1}^2\!+\!n_2(G_{\setminus\alpha}){\delta_{q}}^2}.\nonumber
  \end{eqnarray}
  The lemma is proved.
\myQED

\section{Proof of Theorem~\ref{thm:nontree}} \label{app:nontree}
\emph{Proof:}
We first prove that for each $S\!\in\!\mathcal{S}_G$ there exists $S^\prime\in\!\mathcal{S}^\prime_G(\varphi)=\{S|s_{n,k}\!=\!e^{i\theta_k} , \theta_1\!=\!0, \theta_2,...,\theta_K\!\in\!(-\infty, \infty), \textrm{for}\ e_{n,k}
\in \bar{\varphi},\textrm{ and }s_{n,k}\!=\!e^{i\theta_{n,k}}, \theta_{n,k}\!\in\!(-\infty, \infty), \textrm{for}\ e_{n,k}
\in \varphi\}$ with $A(S^\prime,Z)=A(S,Z)$.
We just need to show that for a given $S\!\in\!\mathcal{S}_G$, there exists an $S^\prime\!\in\!\mathcal{S}^\prime_G$ which is a row rotation of $S$. Assume that $S$ is given. We determine $S^\prime$ as follows. Since the zero elements in $S^\prime$ are predetermined by the factor graph $G$, we only determine the nonzero elements in $S^\prime$. The procedure is similar as that in the proof of Theorem~\ref{thm:tree} except some modifications.

i.  For each $n\in\{n|e_{n,1}\in \bar{\varphi}\}$,  the $n$-th row
of $S^\prime$ is a rotation

\  \ \ of the $n$-th row of $S$, i.e., $s^\prime_{n,k}=s_{n,k}/s_{n,1}$ for $e_{n,k}\in E$.

ii. Find a column of $S^\prime$ that satisfies the following two

\ \ \ conditions. a) The column has a determined labeling for

\ \ \  an edge in $\bar{\varphi}$. b) The column has at least one undetermined

\ \ \  labeling for an edge in $\bar{\varphi}$.  If the
 $m$-th column is found

\ \ \ and $s^\prime_{j,m}$ is the determined
labeling for $e_{j,m}\in\bar{\varphi}$, For each

\ \ \ $n\in\{n|e_{n,m}\in\bar{\varphi}, n\neq j\}$, the $n$-th row of $S^\prime$ is a rotation of

\ \ \  the $n$-th row of $S$, i.e.,$s^\prime_{n,k}=s^\prime_{j,m}s_{n,k}/s_{n,m}$ for $e_{n,k}\in E$.

iii. If all the labelings for the edges in $\bar{\varphi}$ are determined,

\ \ \ terminate the procedure, otherwise, repeat step ii.\\
Note that if there are still undetermined labelings for edges in $\varphi$ at the end of the procedure, we simply use the same labeling
as in $S$.  Moreover,
the above procedure only applies to the case that the remaining graph after deleting edges in $\varphi$ is connected. If it is not connected, i.e., it contains multiple trees, we can label each of them independently in a similar way.

Step ii can always be successful since the remaining graph with edges in $\bar{\varphi}$  is a tree. Step ii guarantees that labeling in each column for the edges in $\bar{\varphi}$ are the same.

Applying the column rotation invariance property of Lemma~\ref{lem:rot2}, we can get a matrix $S^*\in\!\mathcal{S}^*_G(\varphi)=\{S|s_{n,k}\!=\!e^{i\theta_k}, \theta_1\!=\!0, \theta_2,...,\theta_K\!\in\![0, \frac{\pi}{2}), \textrm{for}\ e_{n,k}
\in \bar{\varphi},\textrm{ and }s_{n,k}\!=\!e^{i\theta_{n,k}^\prime}, \theta_{n,k}^\prime\!\in\![0, 2\pi), \textrm{for}\ e_{n,k}
\in \varphi\}$ through column rotations from $S^\prime$ with $A(S^*,Z)=A(S^\prime,Z)=A(S,Z)$. Thus, the theorem is proved.
\myQED


\end{document}